\documentclass[11pt]{article}

\usepackage{amssymb,amsmath,amsfonts,amssymb}
\usepackage{graphics,graphicx,color,psfig}

\hsize \textwidth
\advance \hsize by -\marginparwidth
\evensidemargin \oddsidemargin
\usepackage{amssymb}
\advance\hoffset by 5mm

\def\@abssec#1{\vspace{.05in}\footnotesize \parindent .2in
{\bf #1. }\ignorespaces}
\newtheorem{theorem}{Theorem}[section]
\newtheorem{lemma}[theorem]{Lemma}

\newtheorem{thm}[theorem]{Theorem}

\def \Rm {\mathbb R}

\newcommand{\eps}{\varepsilon}
\newcommand{\E}{\mathbb E}
\newcommand{\dsum}{\displaystyle\sum}
\newcommand{\dint}{\displaystyle\int}
\newcommand{\pdr}[2]{\dfrac{\partial{#1}}{\partial{#2}}}

\newcommand{\bk}{\mathbf k}\newcommand{\vy}{\mathbf y}
\newcommand{\bx}{\mathbf x} \newcommand{\by}{\mathbf y}
\newcommand{\bK}{\mathbf K}\newcommand{\vX}{\mathbf X}
 \newcommand{\vx}{\mathbf x}

\newcommand{\bp}{\mathbf p} \newcommand{\bq}{\mathbf q}

\newcommand{\veta}{{\boldsymbol{\eta}}}
\newcommand{\commentout}[1]{}

\renewcommand{\thefootnote}{{\arabic{footnote}}}

\newcommand{\bxi}{\boldsymbol{\xi}}\newcommand{\vxi}{\boldsymbol{\xi}}

 \renewcommand{\arraystretch}{1.5}

\title{Self-averaging in time reversal for the parabolic wave
  equation}

\author{Guillaume Bal \footnote{Department of 
   Applied Physics and Applied Mathematics, Columbia University, 
   New York NY, 10027; gb2030@columbia.edu}
\and George Papanicolaou 
\footnote{Department of Mathematics, Stanford CA, 94305;
papanico@math.stanford.edu}
\and Leonid Ryzhik \footnote{Department of
Mathematics, University of Chicago, Chicago IL, 60637;
ryzhik@math.uchicago.edu}
}

\begin{document}
 
\maketitle

%tableofcontents

\begin{abstract}
We analyze the self-averaging properties of time-reversed
solutions of the paraxial wave equation with random coefficients, 
which we take to be Markovian in the direction of propagation. This
allows us to construct an approximate martingale for the phase space
Wigner transform of two wave fields. Using a priori $L^2$-bounds
available in the time-reversal setting, we prove that the Wigner
transform in the high frequency limit converges in probability to its
deterministic limit, which is the solution of a transport equation.
\end{abstract}
 
%\begin{AMS}
%\end{AMS}

\renewcommand{\thefootnote}{\fnsymbol{footnote}}
\renewcommand{\thefootnote}{\arabic{footnote}}

\renewcommand{\arraystretch}{1.1}

%\begin{keywords}
%\end{keywords}

%\begin{AMS}
%\end{AMS}

%\pagestyle{myheadings}
%\thispagestyle{plain}

%%%%%%%%%%%%%%%%%%%%%%
%%% BEGINNING TEXT %%%
%%%%%%%%%%%%%%%%%%%%%%

\section{Introduction}
\label{sec:intro}

In time-reversal experiments 
%\cite{Bal-Ryzhik1,Bal-Ryzhik2,Bardos-Fink,BPZ-JASA01,Ewodo,Fouque-Clouet,PRS}
%\cite{DJ90,Fink-PT,Fink-Nonlin,Fink-Prada-01,KHSAFJ}.  
a signal emitted by a
localized source is recorded at an array of
transducers. It is then re-emitted into the medium
reversed in time, that is, the part of the signal that is recorded
first is sent back last.  Because of the time-reversibility of the
wave equation the back-propagated signal refocuses
approximately at the location of the original source 
because the array is limited in size. 
A striking experimental observation is that the presence of
inhomogeneities in the medium improves the refocusing resolution. 
The explanation for this super-resolution
is multipathing: waves in complex media that are
captured by the recording array have undergone
multiple scattering making it effectively larger than
its physical size. 
Another important feature of super-resolution
in time reversal is that the refocused
signal does not depend on the realization of the random medium. 
That is, the refocused signal is deterministic.
Super-resolution and self-averaging of refocused signals in
complicated media has been observed both in laboratory
experiments (see reviews \cite{Fink-PT,FCDPRTTW} and references
therein) and in underwater acoustic wave propagation over long
distances (tens of kilometers) \cite{DJ90,KHSAFJ}.
Time-reversal techniques have numerous applications ranging from
medicine to communications and, more recently, imaging in random media
\cite{BBPT,BPTB,Fink-Prada-01}.
%This paper provides a mathematical explanation of
%these two properties.  

%Finally, the
%refocusing phenomenon is very stable with respect to signal processing
%at the array of recorders, and does not require to send back the exact
%time reversed signal.

The first mathematical analysis of time-reversal in random media was
given by Clouet and Fouque \cite{Fouque-Clouet}, who analyzed
refocusing and self-averaging of time-reversed pulses in a
one-dimensional layered random medium. Their result was extended to a
three-dimensional layered medium in \cite{Ewodo}. Super-resolution in
spatial refocusing and its statistical stability for multi-dimensional
waves in random media was analyzed in \cite{BPZ-JASA01,PRS}, in a
remote-sensing regime where the paraxial or parabolic wave equation
can be used.  The refocusing of the average signal in a full
three-dimensional medium, in the regimes of random geometrical optics
and radiative transfer (transport), was studied in
\cite{Bal-Ryzhik1,Bal-Ryzhik2}. We also mention that another source of
multipathing is the mixing of waves by the boundaries in an ergodic
cavity. This has been studied experimentally in \cite{Fink-Nonlin} and
mathematically in \cite{Bardos-Fink}.

The purpose of this paper is to analyze time reversal in the radiative
transfer regime using the parabolic wave equation, when the waves
interact fully with the random inhomogeneities. We prove mathematically
that the refocused signal is self-averaging, 
which means that it
does not depend on the realization of the random medium. The 
mathematical quantitiy that we analyze
is the Wigner measure of a pair of oscillatory
solutions of the random Schr\"odinger equation. In the present
setting, the random potential depends in a Markovian way on the
variable $z$, the main direction of propagation of the waves.  This
allowed us to use in \cite{BPR-Nonlin} a martingale method to prove
that the average of the Wigner distribution converges to a solution of
the radiative transfer equation. In this paper we use additional
regularity of the Wigner measure, available in time-reversal
when there is some blurring at the recording array, to
show that the whole Wigner distribution, and not only its average,
converges weakly, as a Schwartz distribution and in probability, to
the deterministic solution of the transport equation.  
The blurring at the recording array provides 
a priori bounds for the Wigner transform in $L^2$. These bounds and the
Markovianity of the random potential in the direction of propagation
make the time-reversal problem more tractable mathematically and allow
us to prove in a fairly simple and straightforward manner
self-averaging of the time-reversed signal.

We recall that the Wigner transform is a convenient tool to analyze
high frequency wave propagation in deterministic \cite{GMMP,LP,tartar}
and random media \cite{RPK-WM}. Introduced by Wigner in 1932
\cite{wigner32}, it has been used extensively in the mathematical
literature recently.  Convergence of the average Wigner distribution
to the solution of the radiative transfer equation was first proved by
H. Spohn in \cite{Spohn} for time-independent potentials on small time
intervals. This result was extended to global in time convergence by
L. Erd\"os and H.-T. Yau \cite{Erdos-Yau2}. These proofs involve
infinite Neumann (diagrammatic) expansions for the solution of the
Schr\"odinger equation and are quite involved technically. The
corresponding problem with time-dependent potentials is much simpler
mathematically. It was treated by us in \cite{BPR-Nonlin} in the
Markovian case, and by F. Poupaud and A. Vasseur \cite{PV} in the case
of finite-range time correlations.  In this paper we use the
fact that the Wigner family arising in time-reversal problems is more
regular than the usual one because blurring is added at
the recording array. This provides some additional regularity usually
obtained by considering mixtures of states as, for instance, in
\cite{LP,PV,Spohn}.
%This allows us to prove not only convergence in
%expectation but also in probability to a deterministic
%limit. 
%Moreover, in the time reversal problem the ``mixture of
%states'' arises naturally as a consequence of a convolution the
%blurring kernel $f$.  

The paper is organized as follows: we describe the scaling and obtain
an expression for the back-propagated signal in terms of the Wigner
transform in Section \ref{sec:signal}. The main result and assumptions on the 
random medium are formulated in Section \ref{sec:mainresult}. The proofs
are presented in Section \ref{sec:wigner}.

This work was partially supported by ONR grants N00014-02-1-0088 and
N00014-02-1-0089. G.~Bal was supported in part by NSF grant
DMS-0072008, G. Papanicolaou by grants AFOSR F49620-01-1-0465 and NSF
DMS-9971972, and L. Ryzhik in part by NSF grant DMS-9971742 and the Alfred
P. Sloan Foundation.
 
\section{The back-propagated signal in the parabolic approximation}
\label{sec:signal}

\subsection{The back-propagated signal}

The pressure field $p(z,\bx,t)$ satisfies the scalar wave equation
\begin{equation}
  \label{eq:wave}
\frac{1}{c^2(z,\bx)}\frac{\partial^2p}{\partial t^2}-\Delta p=0.
\end{equation}
%A schematic description of the time reversal procedure for the wave
%equation is presented in Fig.  \ref{fig:onestep}.  
%\begin{figure}[htbp]
%  \begin{center}
%   \hspace{-1cm}\includegraphics[height=4.5cm]{refocusing.eps}
%    \caption{The Time Reversal Procedure. Here,
%   $p_t$ denotes $\pdr{p}{t}$.}
%    \label{fig:onestep}
%  \end{center}
%\end{figure}
Here $c(z,\vx)$ is the local wave speed that we will assume to be
random, and the Laplacian operator includes both direction of
propagation, $z$, and the transverse variable $\vx\in{\mathbb R}^d$.
In the physical setting, we have $d=2$. We consider dimensions
$d\geq1$ to stress that our analysis is independent of the dimension.
If we assume that at time $t=0$, the wave field has a ``beam-like''
structure in the $z$ direction, and if back-scattering may be
neglected, we can replace the wave equation by its parabolic
approximation \cite{Tappert}. More precisely, the pressure $p$ may be
approximated as
\begin{equation}
  \label{eq:ansatz}
p(z,\bx,t)\approx 
\dint_{\Rm} e^{i(-c_0\kappa t+\kappa z)}\psi(z,\bx,\kappa) c_0 d\kappa,
\end{equation}
where $\psi$ satisfies the Schr\"odinger equation
\begin{equation}
\label{eq:sch}
\begin{array}{l}
2i\kappa \pdr{\psi}{z}(z,\bx,\kappa)+\Delta_{\vx}\psi(z,\bx,\kappa)+
\kappa^2 (n^2(z,\bx)-1)\psi(z,\bx,\kappa)=0,\\
  \psi(z=0,\vx,\kappa)=\psi_0(\vx,\kappa)
  \end{array}
\end{equation}
with $\Delta_{\vx}$ the transverse Laplacian in the variable $\vx$.
The refraction index $n(z,\bx)={c_0}/{c(z,\bx)}$, and $c_0$ in
(\ref{eq:ansatz}) is a reference speed.
%We will assume that the
%``initial data'' $\psi_0(\vx,\kappa)$ is smooth and uniformly bounded
%in $\vx$ and has uniformly compact support for all $\kappa$:
%\begin{equation}\label{assump-psi0}
%\|\psi_0(\vx,\kappa)\|_{C^p({\mathbb R}^d)}\le C,~~~
% \psi_0(\vx,\kappa)=0~~\hbox{for $|\vx|\ge l$}
%\end{equation}
%with a sufficiently large $p$.
The rigorous passage to the parabolic approximation from the wave
equation has been analyzed in \cite{parabolic} in the deterministic
case, and \cite{jpf} in a one dimensional random medium. A formal
derivation of the paraxial approximation that leads to the radiative
transfer regime is given below in section \ref{sec:scaling}.

\begin{figure}[htbp]
  \begin{center}
    \includegraphics[height=6cm]{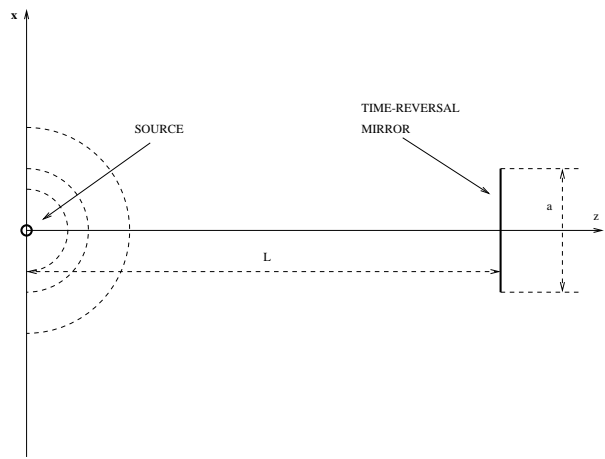}
    \caption{Geometry of the time-reversal experiment.}
    \label{fig1}
  \end{center}
\end{figure}

We assume that the original source is located in the plane $z=0$, and
the time-reversal mirror is located in the plane $z=L$ as depicted in Figure 
\ref{fig1}.  

During the first
stage of a time reversal experiment the signal propagates for a time
$T$, or equivalently, over a distance $L=c_0 T$, so that the signal 
arriving at the time-reversal mirror is given by
\begin{displaymath}
  \psi_-(L,\bx)=\dint_{\Rm^d} G(L,\bx,\kappa;\vy)\psi_0(\vy)d\vy,
\end{displaymath}
where the Green's function solves
\begin{equation}
  \label{eq:schgreen}
  \begin{array}{l}
  2i\kappa \pdr{G}{z}(z,\bx,\kappa;\vy) + 
   \Delta_{\vx} G(z,\bx,\kappa;\vy)+\kappa^2(n^2(z,\vx)-1)
G(z,\bx,\kappa;\vy)=0,\\
  G(0,\vx,\kappa;\vy)=\delta(\vx-\vy).
  \end{array}
\end{equation}
Then the signal is time reversed. For three-dimensional acoustic
pulses, this means that the pressure field is kept unchanged and that
the sign of its time derivative is reversed. In the parabolic
approximation, this is equivalent to phase conjugation, where $\psi$
is replaced by its complex conjugate $\psi^*$. We assume that the
recording array occupies a compact subset of the plane $z=L$, and
introduce a real-valued aperture function $\chi(\bx)$. It represents
the restriction of the signal onto the array, and possible
amplification by the array that may vary from one receiver to
another. In the absence of amplification it is given by the
characteristic function of an array set $\Omega\subset\Rm^d$.  We also
allow for some blurring of the recorded signal, modeled by a
convolution with kernel $f(\bx)$. The signal at $L$ after time
reversal takes then the form
\begin{eqnarray*}
&&  \psi_+(L,\bx,\kappa)=\int_{{\mathbb R}^d}
\chi(\vx)f(\bx-\vy)\psi^*_-(L,\vy;\kappa)\chi(\vy)d\vy\\
&&= \dint_{\Rm^{2d}} \chi(\vx)G^*(L,\by,\kappa;\vy')\chi(\vy)
    f(\bx-\by)\psi_0^*(\vy',\kappa) d\vy d\by'.
\end{eqnarray*}
The last step consists in letting the signal propagate back to the
origin $z=0$ and time reversing it one more time
\begin{equation}
  \label{eq:psiB}
  \psi^B(\bx,\kappa)=\dint_{\Rm^{3d}} G^*(L,\bx,\kappa;\veta)
G(L,\by,\kappa;\by')\chi(\veta)\chi(\by)
    f(\veta-\by)\psi_0(\by',\kappa) d\by d\vy' d\veta.
\end{equation}
The last step, phase-conjugating at the source, is not performed in
real physical experiments but is convenient if we need to
compare the back-propagated signal to the original signal. It affects
neither the degree of refocusing nor the self-averaging effect. The
back-propagated signal in time is given at the plane $z=0$ by
\begin{eqnarray}\label{eq:psiB-time}
&&\psi^B(\vx,t)=\int_{\mathbb R}e^{-ic_0\kappa t}
\psi^B(\bx,\kappa)d\kappa\\
&&~~~~~~~~~~
=\dint_{{\mathbb R}^{3d+1}}\! G^*(L,\bx,\kappa;\veta)e^{-ic_0\kappa t}
G(L,\by,\kappa;\by')\chi(\veta)\chi(\by)
f(\veta-\by)\psi_0(\by',\kappa) d\by d\vy' d\veta d\kappa.\nonumber
\end{eqnarray}
We will be interested in the sequel in the refocusing and
self-averaging properties of the back-propagated signal
$\psi^B(\bx,\kappa)$ for each fixed frequency. Self-averaging of the
time signal $\psi^B(\bx,t)$ then follows from (\ref{eq:psiB-time})
when $\psi_0(\vy,\kappa)$ has compact support in $\kappa$.  An
interesting conclusion of this paper is that the additional blurring
by convolution with the kernel $f$ makes the signal
$\psi^B(\bx,\kappa)$ self-averaging for every frequency $\kappa$. This
should be contrasted with the situation studied in
\cite{BPZ-JASA01}, where no blurring was introduced and self-averaging
was observed only for the full time signal.

\subsection{Localized Source and Scaling}
\label{sec:scaling}

We recall the formal passage from the reduced wave equation to
the parabolic equation (\ref{eq:sch}) described in the Appendix to
\cite{BPR-Nonlin} and explain how the
radiative transfer scaling arises in this context. We start with the
reduced wave equation
\begin{equation}\label{red-wave}
\Delta \hat p+\kappa^2n^2(z,\vx)\hat p=0,
\end{equation}
and look for solutions of (\ref{red-wave}) in the form
$
\hat p(z,\vx)=e^{i\kappa z}\psi(z,\vx).
$
Then we obtain
\begin{equation}\label{red-unscaled}
\frac{\partial^2\psi}{\partial z^2}+2i\kappa\pdr{\psi}{z}+\Delta_\vx\psi+
\kappa^2(n^2-1)\psi=0.
 \end{equation}
The refraction index $n(z,\vx)$ is weakly fluctuating so that it has the form
\[
n^2(z,\vx)=1-2\sigma V\left(\frac{z}{l_z},\frac{\vx}{l_x}\right),
\]
where $V$ models random fluctuations that will be described in detail
in section \ref{sec:random-refr-ind}.  Here, $l_x$ and $l_z$ are the
correlation lengths of $V$ in the transverse and longitudinal
directions, respectively, and the small parameter $\sigma$ measures
the strength of the fluctuation. The waves propagate over a distance
$L_x$ in the $\vx$-plane and $L_z$ in the $z$-direction, and we
rescale $\vx$ and $z$ accordingly. We also introduce a carrier wave
number $\kappa_0$ and replace $\kappa=\kappa_0\kappa'$, $\kappa'$
being the non-dimensional wave number (we drop the prime below).  The
physical parameters determined by the medium are the length scales
$l_x$, $l_z$ and the non-dimensional parameter $\sigma\ll 1$. 

We now explain the scaling of the parameters $L_x$, $L_z$ and
$\kappa_0$ in the radiative transfer regime.  Equation
(\ref{red-unscaled}) in the non-dimensional variables $z'=z/L_z$,
$\vx'=\vx/L_x$ becomes after we drop the primes
\begin{equation}\label{red-scaled}
\frac{1}{L_z^2}\frac{\partial^2\psi}{\partial z^2}+
\frac{2i\kappa\kappa_0}{L_z}\pdr{\psi}{z}+
\frac{1}{L_x^2}\Delta_\vx\psi-
2\kappa^2\kappa_0^2
\sigma V\left(\frac{zL_z}{l_z},\frac{\vx L_x}{l_x},\right)\psi=0.
\end{equation}
We introduce the following small parameters:
\[
\delta_x=\frac{l_x}{L_x},~~\delta_z=\frac{l_z}{L_z},~~~
\gamma_x=\frac{1}{\kappa_0l_x},~~\gamma_z=\frac{1}{\kappa_0l_z}
\]
and rewrite (\ref{red-scaled}) as
\begin{equation}\label{red-param}
\gamma_z\delta_z\frac{\partial^2\psi}{\partial z^2}+
2i\kappa\pdr{\psi}{z}+
\frac{\delta_x^2\gamma_x^2}{\delta_z\gamma_z}\Delta_\vx\psi-
\frac{2\kappa^2\sigma}{\gamma_z\delta_z}
V\left(\frac{z}{\delta_z},\frac{\vx}{\delta_x},\right)\psi=0.
\end{equation}
Let us assume now that 
\begin{equation}\label{parab-assump}
\delta_x=\delta_z\ll 1,~~~\gamma_z=\gamma_x^2\ll 1,~~
\sigma=\gamma_z\sqrt{\delta_x}
\end{equation}
and denote $\eps=\delta_x$. Then (\ref{red-param}) becomes
after multiplication by $\eps/2$
\begin{equation}\label{red-param1}
\frac{\gamma_z\eps^2}{2}\frac{\partial^2\psi}{\partial z^2}+
i\kappa\eps\pdr{\psi}{z}+
\frac{\eps^2}{2}\Delta_\vx\psi-\kappa^2\sqrt{\eps}
V\left(\frac{z}{\eps},\frac{\vx}{\eps}\right)\psi=0.
\end{equation}
Observe that when $\kappa=O(1)$ and $\gamma_z\ll 1$, the first term in
(\ref{red-param1}) is small and may be neglected in the leading order
since $|\eps^2\psi_{zz}|=O(1)$. Then (\ref{red-param1}) becomes
\begin{equation}\label{eq:sch-rescaled}
i\kappa\eps\pdr{\psi}{z}+
\frac{\eps^2}{2}\Delta_\vx\psi-\kappa^2\sqrt{\eps}
V\left(\frac{z}{\eps},\frac{\vx}{\eps}\right)\psi=0
\end{equation}
which is the parabolic wave equation (\ref{eq:sch}) in the radiative
transfer scaling.

The second condition in (\ref{parab-assump}) implies that the carrier
wave number should be chosen as $\kappa_0=l_z/l_x^2$. Then
$\gamma_z=l_x^2/l_z^2\ll 1$ implies that $l_x\ll l_z$.  Therefore the
correlation length in the longitudinal direction $z$ should be much
longer than in the transverse plane $\bx$. Furthermore, we should have
$\eps=(\sigma/\gamma_z)^2=\sigma^2l_z^4/l_x^4\ll 1$ which in turn
implies that the fluctuation strength $\sigma\ll (l_x/l_z)^4$. Given
that these constraints are satisfied the last condition in
(\ref{parab-assump}) implies that the spatial scales $L_x$ and $L_z$
should be chosen according to
\begin{displaymath}
  L_x=l_x\frac{l_x^4}{\sigma^2l_z^4},\qquad
  L_z=l_z\frac{l_x^4}{\sigma^2l_z^4}.
\end{displaymath}
Note that the constraint $l_x\ll l_z$ implies that $L_x\ll L_z$, which
is the usual constraint for the validity of the parabolic
approximation.  The connection between the relations of the physical
parameters and the resulting scaling of the parabolic approximation is
further discussed in \cite{PRS}. The rigorous passage to the parabolic
approximation in one dimension in a random medium in a similar scaling
is discussed in
\cite{jpf}.

To define what we mean by the quality of the refocused signal, we
assume that the initial pulse is centered around a point $\vx_0$
with support comparable to the transverse correlation length. Then the
initial data for the Schr\"odinger equation (\ref{eq:sch-rescaled})
becomes in the rescaled variables
\begin{equation}
\label{indata-rescaled}
\psi(z=0,\vx,\kappa)=\psi_0(\frac{\vx-\vx_0}{\eps},\kappa).
\end{equation}
The transducers should obviously be capable of capturing
signals of frequency $\eps^{-1}$ so that the blurring occurs on
the scale of the source. We therefore replace $f(\bx)$ by
$\eps^{-d}f(\bx/\eps)$.  Finally, we are interested in the
refocusing properties of $\psi^B(\bx)$ in the vicinity of $\bx_0$ and
consequently introduce the scaling $\bx=\bx_0+\eps\bxi$. We then
recast (\ref{eq:psiB}) as
\begin{equation}
\label{eq:psiBeps}
\psi^B_\eps(\bxi,\kappa;\bx_0)=\dint_{\Rm^{3d}}
G^*(L,\bx_0+\eps\bxi,\kappa;\veta)
   G(L,\by,\kappa;\bx_0+\eps\by') \chi(\veta,\by) \psi_0(\vy',\kappa)
   d\by' d\by d\veta,
\end{equation}
where 
\begin{equation}\label{chieps}
 \chi(\veta,\by)= \chi(\veta)\chi(\by)f(\dfrac{\veta-\by}{\eps}).
\end{equation}
We observe that
$G(L,\bx,\kappa;\by)=G(L,\by,\kappa;\bx)$,
so that
\begin{equation}
  \label{eq:psiBeps2}
\psi^B_\eps(\bxi,\kappa;\bx_0)=\dint_{\Rm^{3d}}
   G^*(L,\bx_0+\eps\bxi,\kappa;\veta)
   G(L,\bx_0+\eps\by',\kappa;\by) \chi(\veta,\by) \psi_0(\vy',\kappa)
   d\by' d\by d\veta.
\end{equation}

Let us now introduce the auxiliary function $Q_\eps$ as in
\cite{Bal-Ryzhik1,Bal-Ryzhik2},
\begin{equation}
  \label{eq:Qeps}
  Q_\eps(L,\bx,\kappa;\bq)=\dint_{\Rm^d} G(L,\bx,\kappa;\by)
    \chi(\by)e^{-i\bq\cdot\by/\eps} d\by,
\end{equation}
which satisfies the initial value problem
\begin{eqnarray}
\label{eq:sch-Q}
&&
i\eps\kappa \pdr{Q_\eps}{z}(z,\bx;\kappa)+\frac{\eps^2}{2}
\Delta_{\vx}Q_\eps(z,\bx;\kappa)-
\kappa^2\sqrt{\eps}
V\left(\frac{z}{\eps},\frac{\bx}{\eps}\right)Q_\eps(z,\bx,\kappa)=0,\\
&&Q_\eps(z=0,\vx;\kappa)=\chi(\vx)e^{-i\bq\cdot\vx/\eps}.\nonumber
\end{eqnarray}
It physically describes the component with wave vector $\bq$ that is
sent back to the plane $z=0$ by the array.  We then define the Wigner
transform as 
\begin{equation}
  \label{eq:Weps-def}
W_\eps(L,\bx,\bk,\kappa)=\dint_{\Rm^d}
\hat f(\bq)U_\eps(L,\bx,\bk,\kappa;\bq)d\bq,
\end{equation}
where
\begin{equation}
  \label{eq:Ueps}
  U_\eps(L,\bx,\bk,\kappa;\bq)= \dint_{\Rm^d}
    e^{i\bk\cdot\by} Q_\eps(L,\bx-\dfrac{\eps\by}{2},\kappa;\bq)
     Q^*_\eps(L,\bx+\dfrac{\eps\by}{2},\kappa;\bq)
     \dfrac{d\by}{(2\pi)^d}.
\end{equation}
The Wigner transform $W_\eps$ is more regular than $U_\eps$ because of
additional averaging in the wave vector $\bq$. More, precisely, while
the family $U_\eps$ is not uniformly bounded in $L^2$, the family
$W_\eps$ has a uniform bound in $L^2$. This regularizing effect is
essentially the same as the one obtained by considering mixtures of
states for the Schr\"odinger equation as in \cite{LP,PV,Spohn}.

We can then recast (\ref{eq:psiBeps2}) as 
\begin{equation}
  \label{eq:psiBeps3}
  \psi^B_\eps(\bxi,\kappa;\bx_0)=\dint_{\Rm^{2d}} 
    e^{i\bk\cdot(\bxi-\by)} 
     W_\eps(L,\bx_0+\eps\dfrac{\vy+\bxi}{2},\bk,\kappa) \psi_0(\vy,\kappa)
    \dfrac{d\vy d\bk}{(2\pi)^d}.
\end{equation}
The above formula shows that the asymptotic behavior of
$\psi^B_\eps(\bxi,\kappa;\bx_0)$ as $\eps\to0$ is characterized by
that of the Wigner transform
$W_\eps(L,\bx,\bk,\kappa)$.

\section{The main results and assumptions on the random medium}
\label{sec:mainresult}
\subsection{The main results}\label{sec:main}

This section presents our two main results. The first one describes
the self-averaging of the back-propagated signal $\psi^B$ in
(\ref{eq:psiBeps3}). Its proof is based on the second theorem, of
independent interest, which shows that the Wigner transform converges
in probability to a unique deterministic limit, solution of a
transport equation.

\begin{thm}\label{thm1}
Let the array function $\chi(\vy)$ and the radially symmetric filter
$f(\vy)$ be in $L^1\cap L^\infty ({\mathbb R}^d)$, while $\psi_0\in
L^2({\mathbb R}^d)$ for a given $\kappa\in{\mathbb R}$. Assume also
that the refraction index $n(z,\vx)$ satisfies assumptions in section
\ref{sec:random-refr-ind} below.  Then for each $\vxi\in{\mathbb R}^d$
the back-propagated signal $\psi_\eps^B(\vxi,\vx_0,\kappa)$ given by
(\ref{eq:psiBeps3}) converges in probability and weakly in
$L^2_{\vx_0}({\mathbb R}^d)$ as $\eps\to 0$ to the deterministic
function
\begin{equation}
  \label{eq:psiB-limit}
\psi^B(\bxi,\kappa;\bx_0)=\dint_{\Rm^{2d}} e^{i\bk\cdot(\bxi-\by)} 
\overline W(L,\bx_0,\bk,\kappa) \psi_0(\vy,\kappa)
\dfrac{d\vy d\bk}{(2\pi)^d}.
\end{equation}
The function $\overline W$ satisfies the transport equation
\begin{equation}
  \label{eq:treqformal}
  \pdr{\overline W}{z}+\frac{1}{\kappa}\bk\cdot\nabla_\bx \overline W=
  \kappa{\cal L}\overline W,
\end{equation}
with initial data $\overline W_0(\vx,\bk)=\hat f(\bk)|\chi(\vx)|^2$
and where the operator $\cal L$ is defined by
\begin{equation}\label{eq:cal-L}
{{\cal L}}\lambda=\int_{\Rm^d} \frac{d\bp}{(2\pi)^d}
\hat R(\frac{|\bp|^2-|\bk|^2}{2},\bp-\bk)(\lambda(\bp)-\lambda(\bk)).
\end{equation}
Here $\hat R(\omega,\bp)$ is the
Fourier transform of the correlation function of $V$, defined by
(\ref{eq:tildeR}) below.
\end{thm}
The convergence of the Wigner transform is described by the following
theorem.
\begin{thm}\label{thm2}
Under the assumptions of Theorem \ref{thm1} the Wigner distribution
$W_\eps$ converges in probability and weakly in $L^2({\mathbb
R}^{2d})$ to the solution $\overline W$ of the
transport equation (\ref{eq:treqformal}). More precisely, for any test
function $\lambda\in L^2({\mathbb R}^{2d})$ the
process $\langle W_\eps(z),\lambda\rangle$ converges to $\langle
\overline W(z),\lambda\rangle$ in probability as $\eps\to 0$,
uniformly on finite intervals $0\le z\le L$. 
\end{thm}           
Here, $\langle \cdot,\cdot\rangle$ is the usual scalar product in
$L^2(\Rm^{2d})$.  Theorem \ref{thm1} is then a corollary of Theorem
\ref{thm2} and Lemma \ref{lemma2} below. 

Theorem \ref{thm1} provides a mathematical explanation for the
important properties of the time-reversal experiment. First, it
ensures that the back-propagated signal is essentially independent of
the realization of the random medium in the high-frequency limit since
$\psi^B(\bxi,\kappa;\bx_0)$ is deterministic. Second, it provides a
quantitative description of the back-propagated field, which may be
written as
\[
\psi^B(\cdot,\kappa;\bx_0)=F(\cdot,\kappa;\bx_0)
   \,{{\star}}\,\psi_0(\cdot,\kappa).
\]
The kernel $F(\bxi,\kappa;\bx_0)$ is the Fourier transform in
$\bk\to\bxi$ of the solution of (\ref{eq:treqformal}). We deduce from
this formula that the back-propagated signal has much better
refocusing properties in random media than in homogeneous media. As is
explained in
\cite{Bal-Ryzhik2}, the reason is that the limiting Wigner measure
$\bar W$ is more regular in its $\bk$ variable when the right-hand
side in (\ref{eq:treqformal}) is present. This leads to better spatial
decay of $F$ and hence $\psi_B$ is localized tighter in a random
medium than in a homogeneous medium.  The improvement of the
refocusing in random media has been carefully analyzed theoretically
and numerically in \cite{BPZ-JASA01}, in the Fokker-Plank
approximation of (\ref{eq:treqformal}).

\subsection{Outline of the proof}

We summarize here the main steps of the derivation of the results
stated in Theorems \ref{thm1} and \ref{thm2}.  We first obtain in
Lemma \ref{lemma1} uniform bounds in $L^2$ for the Wigner transform
$W_\eps$ independently of the realization of the random medium. Lemma
\ref{lemma2} then shows that Theorem \ref{thm1} is a consequence of
Theorem
\ref{thm2}. 

The main mathematical restriction is to assume that the random
potential $V(z,\bx)$ is Markovian in its first variable. We refer to
section \ref{sec:random-refr-ind} for the details of its construction.
Following our previous work in \cite{BPR-Nonlin}, we show the
convergence of the expectation $\E\left\{\langle
W_\eps,\lambda\rangle\right\}\to\langle\overline W,\lambda\rangle$ for
every test function $\lambda$. Its proof is simpler (and less general)
than in \cite{BPR-Nonlin} because of the available a priori $L^2$
bounds on the Wigner transform, which satisfies the following Cauchy
problem
\begin{eqnarray}
  \label{eq:Weps}
&&\pdr{W_\eps}{z} + \bk\cdot\nabla_{\bx} W_\eps
={\cal L}_\eps W_\eps\\
&&   W_\eps(0,\bx,\bk)=W^0_\eps(\bx,\bk),\nonumber
\end{eqnarray}
with
\[
{\cal L}_\eps W_\eps=\dfrac{1}{i\sqrt\eps}\dint_{\Rm^d}
\dfrac{d\tilde V(\dfrac{z}{\eps},\bp)}{(2\pi)^d}e^{i{\bp\cdot\bx}/{\eps}}
\left[W_\eps(\bx,\bk-\dfrac{\bp}{2})-W_\eps(\bx,\bk+\dfrac{\bp}{2})\right].
\]
Since $\kappa$ is a fixed parameter, we set $\kappa=1$ without loss of
generality.  The solution of (\ref{eq:Weps}) with initial data in
$L^2$ is understood in the sense that for every smooth test function
$\lambda(z,\vx,\bk)$, we have
\[
\langle W_\eps(z),\lambda(z)\rangle-\langle W_\eps^0,\lambda(0)\rangle
=\int_0^z\langle W_\eps(s),
\left(\pdr{}{s}+\bk\cdot\nabla_\vx+
{\cal L}_\eps\right)\lambda(s)\rangle ds.
\]
Here, we have used that ${\cal L}_\eps$ is a
self-adjoint operator for $\langle \cdot,\cdot\rangle$.  Therefore for
a smooth function $\lambda_0(\vx,\bk)$ we obtain $\langle
W_\eps(z),\lambda_0\rangle=\langle W_\eps^0,\lambda_\eps(0)\rangle$,
where $\lambda_\eps(s)$ is the solution of the backward problem
\[
\pdr{\lambda_\eps}{s}+\bk\cdot\nabla_\vx\lambda_\eps+
{\cal L}_\eps\lambda_\eps(s)=0,~~
0\le s\le z
\]
with the terminal condition $\lambda_\eps(z,\vx,\bk)=\lambda_0(\vx,\bk)$.
This defines the process $W_\eps(z)$ in $L^2({\mathbb R}^{2d})$ and
generates a corresponding measure $P_\eps$ on the space
$C([0,L];L^2({\mathbb R}^{2d}))$ of continuous functions in time with
values in $L^2$. The measure $P_\eps$ is actually supported on paths
inside a ball $X=\left\{W\in L^2:~\|W\|_{L^2}\le C\right\}$ with the
constant $C$ as in Lemma \ref{lemma1}. The set $X$ is the state space
for the random process $W_\eps(z)$. We have proved in
\cite{BPR-Nonlin} and will use in the sequel that the family $P_\eps$
is tight:
\begin{lemma}\label{lemma-tight}
The family of measures $P_\eps$ is weakly compact.
\end{lemma}

The proof of convergence of $W_\eps$ to its deterministic limit is
obtained in two steps.  Let us fix a deterministic test function
$\lambda(z,\vx,\bk)$.  We use the Markovian property of the random
field $V(z,\vx)$ in $z$ to construct a first functional
$G_\lambda\!:C([0,L];X)\!\to C[0,L]$ by
\begin{equation}\label{G-lambda}
G_\lambda[W](z)=\langle W,\lambda\rangle(z)-\int_0^z\langle
W,\pdr{\lambda}{z}+\bk\cdot\nabla_\vx\lambda+ {\cal
L}\lambda\rangle(\zeta)d\zeta
\end{equation}
and show that it is an approximate $P_\eps$-martingale. More
precisely, we show that
\begin{equation}\label{eq:approxmart}
\left|\E^{P_\eps}\left\{G_\lambda[W](z)|{\cal F}_s\right\}
-G_\lambda[W](s)\right|\le
C_{\lambda,L}\sqrt{\eps}
\end{equation}
uniformly for all $W\in C([0,L];X)$ and $0\le s<z\le L$. 
Lemma \ref{lemma-tight} implies that there exists a subsequence
$\eps_j\to 0$ so that $P_{\eps_j}$ converges weakly to a measure $P$
supported on $C([0,L];X)$. Weak convergence of $P_\eps$ and the strong
convergence (\ref{eq:approxmart}) together imply that
$G_\lambda[W](z)$ is a $P$-martingale so that
\begin{equation}\label{eq:Pmart}
\E^{P}\left\{G_\lambda[W](z)|{\cal F}_s\right\}-G_\lambda[W](s)=0.
\end{equation}
Taking $s=0$ above we obtain as in \cite{BPR-Nonlin} the transport
equation (\ref{eq:treqformal}) for $\overline
W=\E^P\left\{W(z)\right\}$ in its weak formulation.

The second step is to show that for every test function
$\lambda(z,\vx,\bk)$ the new functional
\begin{eqnarray*}
&&G_{2,\lambda}[W](z)=\langle W,\lambda\rangle^2(z)-
2\int_0^z\langle W,\lambda\rangle(\zeta)
\langle W,\pdr{\lambda}{z}+\bk\cdot\nabla_\vx\lambda+
{\cal L}\lambda\rangle(\zeta)d\zeta
\end{eqnarray*}
is also an approximate $P_\eps$-martingale. We then obtain that
$\E^{P_\eps}\left\{\langle
W,\lambda\rangle^2\right\}\to\langle\overline W,\lambda\rangle^2$,
which implies convergence in probability. It follows that the limit
measure $P$ is unique and deterministic, and that the whole sequence
$P_\eps$ converges.

That $G_{2,\lambda}[W](z)$ is an approximate $P_\eps$-martingale uses
very explicitly the uniform a priori $L^2$ bound on the Wigner
distribution $W_\eps$. When the a priori bound is available only in a
much larger space (the space ${\cal A}'$ of \cite{BPR-Nonlin}), we are
not able to prove that the functional $G_{2,\lambda}$ is an
approximate $P_\eps$-martingale, and actually suspect that the result
is not true.  We expect to observe that some long-range correlations
survive for sufficiently singular initial data, which would imply that
the limiting measure $P$ is no longer deterministic.

\subsection{The random refraction index}\label{sec:random-refr-ind}

We describe here the construction of the random potential $V(z,\vx)$.
The refraction index $n(z,\vx)$ in non-dimensional variables is a
random function of the form
\[
n^2(z,\vx)=1-2\sigma V(z,\vx),
\]
where the non-dimensional parameter $\sigma$ measures the strength of
fluctuations.

The mathematical analysis of the Wigner distribution in random media
can be obtained by the method of diagrammatic expansions of the
solution of the Schr\"odinger equation as in
\cite{Erdos-Yau2,Spohn}. However, here we assume that the random field
$V(z,\vx)$ is a Markov process in $z$, as in \cite{BPR-Nonlin}, and
we are able to analyze the Wigner transform $W_\eps$ directly. The
Markovian hypothesis is crucial to simplify the mathematical analysis
because it allows us to treat the process
$z\mapsto(V(z/\eps,\vx/\eps),W_\eps(z,\bx,\bk))$ as jointly Markov.

In addition, $V(z,\vx)$ is assumed to be stationary in $\vx$ and $z$,
mean zero, and is constructed in the Fourier space as follows. Let
${\cal V}$ be the set of measures of bounded total variation with
support inside a ball $B_L=\left\{|\bp|\le L\right\}$
\begin{equation}\label{defcalv}
{\cal V}=\left\{\hat V:~\int_{{\mathbb R}^d}|d\hat V|\le C,
~\hbox{supp}~{\hat
V}\subset B_L,~ \hbox{$\hat V(\bp)=\hat V^*(-\bp)$}\right\}
\end{equation}
and let $\tilde V(z)$ be a mean-zero Markov process on $\cal V$
with generator $Q$. The random potential $V(z,\vx)$ is given by
\[
V(z,\vx)=\int_{\Rm^d} \frac{d\tilde V(z,\bp)}{(2\pi)^d}e^{i\bp\cdot \vx}
\]
and is real and uniformly bounded:
\[
|V(z,\vx)|\le C.
\]
We assume that the process $V(z,\vx)$ is stationary in $z$ and $\vx$
with correlation function $R(z,\vx)$
\[
\E\left\{V(s,\vy)V(z+s,\vx+\vy)\right\}=R(z,\vx)~~~\hbox{for all
$\vx,\vy\in{\mathbb R}^d$, and $z,s\in{\mathbb R}$.}
\]
In terms of the process $\tilde V(z,\bp)$ this means that given any two
bounded continuous functions $\hat\phi(\bp)$ and $\hat\psi(\bp)$ we have
\begin{equation}\label{powerspectrum}
\E\left\{\langle \tilde V(s),\hat\phi\rangle\langle
\tilde V(z+s),\hat\psi\rangle\right\}=
(2\pi)^d\int_{\Rm^d} d\bp \tilde R(z,\bp)\phi(\bp)\hat\psi(-\bp).
\end{equation}
Here $\langle\cdot,\cdot\rangle$ is the usual duality product on
${\mathbb R}^{d}\times\Rm^d$, and the power spectrum $\tilde R$ is the
Fourier transform of $R(z,\vx)$ in $\vx$:
\[
\tilde R(z,\bp)=\int_{\Rm^d} d\bx  e^{-i\bp\cdot \vx}R(z,\vx) .
\]
We assume that $\tilde R(z,\bp)\in {\cal S}({\mathbb
R}\times{\mathbb R}^d)$ for simplicity and define $\hat R(\omega,\bp)$
as
\begin{equation}\label{eq:tildeR}
\hat R(\omega,\bp)=\int_{\Rm}  dz e^{-i\omega z}\tilde R(z,\bp),
\end{equation}
which is the space-time Fourier transform of $R$.
 
We assume that the generator $Q$ is a bounded operator on
$L^\infty({\cal V})$ with a unique invariant measure $\pi(\hat V)$
\[
Q^*\pi=0.
\]
and that there exists $\alpha>0$ such that if $\langle g,\pi\rangle=0$
then
\begin{equation}\label{eq:expdecaynew}
\|e^{rQ}g\|_{L_{\cal V}^\infty}\le C\|g\|_{L_{\cal
V}^\infty}e^{-\alpha r}.
\end{equation}
The simplest example of a generator with gap in the spectrum and
invariant measure $\pi$ is a jump process on ${\cal V}$ where
\[
Qg(\hat V)=\int_{\cal V}g (\hat V_1)d\pi(\hat V_1)-g(\hat V),~~~
\int_{\cal V}d\pi(\hat V)=1.
\]
Given (\ref{eq:expdecaynew}), the Fredholm alternative holds for the
Poisson equation
\[
Qf=g,
\]
provided that $g$ satisfies $\langle\pi,g\rangle=0$.  It has a unique
solution $f$ with $\langle\pi,f\rangle=0$ and $\|f\|_{L^\infty_{V}}\le
C\|g\|_{L^\infty_{V}}$. The solution $f$ is given explicitly by
\[
f(\hat V)=-\int_0^\infty dr e^{rQ}g(\hat V),
\]
and the integral converges absolutely because of
(\ref{eq:expdecaynew}).

The particular Markovian model adopted in this paper is somewhat
restrictive.  However, the only information about the process that
enters into the main result, the transport equation for the Wigner
distribution in the limit $\eps\to 0$, is the two-point correlation
function of $V$. Therefore, we expect that self-averaging of the
Wigner transform and convergence to the solution of the transport
equation hold for much more general classes of random perturbations of
the refraction index.

\section{Convergence of the Wigner transform}
\label{sec:wigner}

\subsection{A priori bounds on the Wigner transform}

The Wigner transform satisfies the following uniform bound.
\begin{lemma}\label{lemma1} 
There exists a constant $C$ that is independent of $\eps$ so that
\begin{equation}\label{weps-bd}
\|W_\eps(z)\|_{L^2({\mathbb R}^{2d})}\le 
C\|f\|_{L^2({\mathbb R}^d)}^2
\|\chi\|_{L^4({\mathbb R}^d)}^4
\end{equation}
for all $z\ge 0$.
\end{lemma}
{\bf Proof.}
It is shown in \cite{LP} that equation (\ref{eq:Weps}) preserves the
$L^2$-norm of $W_\eps$:
$\|W_\eps(z)\|_{L^2}=\|W_\eps(z=0)\|_{L^2}$. The initial conditions
for $W_\eps$ are given by
\[
W_\eps(0,\bx,\bk)=\int_{\Rm^{2d}} \frac{d\by d\bq}{(2\pi)^{d}}
  e^{i\bk\cdot\vy+i\bq\cdot\vy}
\hat f(\bq)\chi(\bx-\dfrac{\eps\by}{2})
\chi(\bx+\dfrac{\eps\by}{2})=
\int_{\Rm^d}d\vy  e^{-i\bk\cdot\vy}f(\vy)\chi(\bx+\dfrac{\eps\by}{2})
\chi(\bx-\dfrac{\eps\by}{2})
\]
so that
\begin{eqnarray*}
&&\int_{\Rm^{2d}} d\vx d\bk  |W_\eps(0,\vx,\bk)|^2\!=\!\int_{\Rm^{4d}}
   \! d\by d\by_1 d\vx d\bk
e^{i\bk\cdot\vy_1-i\bk\cdot\vy}
f(\by)\overline{f(\by_1)}
\chi(\bx-\dfrac{\eps\by}{2}) \chi(\bx+\dfrac{\eps\by}{2})\\
&&\times\chi(\bx-\dfrac{\eps\by_1}{2})
\chi(\bx+\dfrac{\eps\by_1}{2})=(2\pi)^d\int_{\Rm^{2d}} d\vx d\vy
|f(\by)|^2
\left|\chi(\bx-\dfrac{\eps\by}{2})\right|^2
\left|\chi(\bx+\dfrac{\eps\by}{2})\right|^2
\end{eqnarray*}
and (\ref{weps-bd}) follows.

The above calculation also shows that the initial Wigner distribution
$W_\eps^0(\vx,\bk)$ converges to
\[
W_0(\vx,\bk)=\hat f(\bk)|\chi(\vx)|^2.
\]
Notice however that one cannot expect strong convergence in $L^2$ of
$W_\eps(z)$ to $\overline W(z)$, the solution of the transport
equation (\ref{eq:treqformal}), because the latter has an $L^2$-norm
that decreases as $z$ increases while the $L^2$-norm of $W_\eps$ is
preserved.

The next Lemma shows that one may drop the term $\eps(\vy+\vxi)/2$ in
the argument of $W_\eps$ in expression (\ref{eq:psiBeps3}) for the
back-propagated signal.
\begin{lemma}\label{lemma2}
Let $\phi(\bx_0)\in L^2({\mathbb R}^d)$, then
\begin{equation}
\left|\dint_{\Rm^d}  d\bx_0 \psi_\eps^B(\bxi;\bx_0) \phi^*(\bx_0)
- \dint_{\Rm^{2d}} d\vx_0d\bk e^{i\bk\cdot\bxi} W_\eps(L,\bx_0,\bk)
\phi^*(\bx_0) \hat\psi_0(\bk) \right|\to 0~~~\hbox{as $\eps\to 0$}
\end{equation}
for all $\vxi$.
\end{lemma}
{\bf Proof.}
We have 
\begin{eqnarray*}
%\label{eq:mts}
&&\dint_{{\mathbb R}^d}\psi_\eps^B(\bxi;\bx_0)
\phi^*(\bx_0)d\bx_0-
\dint_{\Rm^{2d}} e^{i\bk\cdot\bxi} W_\eps(L,\bx_0,\bk)\phi(\bx_0)
\hat\psi_0(\bk) d\vx_0d\bk\\
&&=\dint_{{\mathbb R}^{2d}}\hat\phi^*(\bq) 
\hat W_\eps(L,\bq,\vy-\vxi) \psi_0(\by)
\left(e^{-i\eps\bq\cdot(\vy+\vxi)/2}-1\right)\dfrac{d\by d\bq}{(2\pi)^d}
\end{eqnarray*}
while
\begin{eqnarray*}
\dint_{{\mathbb R}^{2d}}|\hat\phi(\bq)|^2|\psi_0(\vy)|^2
\left|1-e^{-i\eps\bq\cdot(\vy+\vxi)/2}\right|^2d\by d\bq
%\le
%C\dint_{{\mathbb R}^2}\!\!d\bq~|\hat\phi(\bq)|^2|\!\!\!\!
%\int\limits_{|\vy|\le l+K}\!\!\!\! d\vy
%\left|1-e^{-i\eps\bq\cdot\vy/2}\right|^2
\to 0
\end{eqnarray*}
by the Lebesgue dominated convergence theorem. Therefore Lemma
\ref{lemma2} follows from Lemma \ref{lemma1}.

%For $\bxi'$ and $\kappa$ given in bounded sets, let us define the test
%function
%\begin{equation}
%  \label{eq:lambda}
%  \lambda(\bx',\bk';\bxi',\kappa)= e^{i\bk'\cdot\bxi'}
%        \phi(\bx_0')\hat\psi_0(\bk',\kappa).
%\end{equation}
%We are interested in the convergence properties (as $\eps\to0$) of
%\begin{equation}
%  \label{eq:mts}
%  \aver{W_\eps,\lambda}=\dint_{\Rm^4}W_\eps(L,\bx',\bk',\kappa)
%   \lambda(\bx',\bk';\bxi',\kappa) d\bx' d\bk'.
%\end{equation}

%Since $W_\eps$ is bounded in ${\cal C}([0,T];L^2(\Rm^4))$, it
%converges weak-* to some function $W(z,\bx,\bk)$ in the same space for
%all fixed $\kappa$ and all realizations of the random potential
%$V(\bx)$. The next Theorem shows that $W_\eps$ converges to $W$ in
%probability.
%\begin{thm}\label{thm2}
%  Let the random filed $V(z,\vx)$ satisfy the assumptions in Section
%  \ref{sec:random-refr-ind}. Let $W_\eps$ be the solution of
%  (\ref{eq:Weps}) with initial data $W_\eps^0(\vx,\bk)$, which
%  converges to $W_0(\vx,\bk)$ weakly in $L^2({\mathbb R}^4)$. Let
%  $\overline W(z,\vx,\bk)$ be the solution of the transport equation
%  (\ref{eq:treqformal}) with initial data $W_0(x,k)$.  Then
%  $W_\eps(z,\vx,\bk)$ converges in probability and weakly in
%  $L^2({\mathbb R}^4)$ to $\overline W(z,\vx,\bk)$, uniformly on
%  finite time intervals.
%\end{thm}               
Theorem \ref{thm1} then follows from Lemma \ref{lemma2} and Theorem
\ref{thm2}, which is proved in the following sections.

\subsection{Convergence of the expectation}\label{sec:expect}

To obtain the approximate martingale property (\ref{eq:approxmart}),
one has to consider the conditional expectation of functionals
$F(W,\hat V)$ with respect to the probability measure $\tilde P_\eps$
on the space $C([0,L];{\cal V}\times X)$ generated by $V(z/\eps)$ and
the Cauchy problem (\ref{eq:Weps}). The only functions we need to
consider are actually of the form $F(W,\hat V)=\langle W,\lambda(\hat
V)\rangle$ with $\lambda\in L^\infty({\cal V};C^1([0,L];{\cal
S}({\mathbb R}^{2d})))$. Given a function $F(W,\hat V)$ let us define
the conditional expectation
\[
\E_{W,\hat V,z}^{\tilde P_\eps}\left\{F(W,\hat V)\right\}(\tau)=
\E^{\tilde P_\eps}\left\{F(W(\tau),\tilde V(\tau))|~W(z)=
W, \tilde V(z)=\hat V\right\},~~ \tau\ge z.
\]
The weak form of the infinitesimal generator of the Markov process
generated by $\tilde P_\eps$ is given by
\begin{equation}
\label{generator}
\left.\frac{d}{dh}\E_{W,\hat V,z}^{\tilde P_\eps}\left\{\langle
W,\lambda(\hat V)\rangle\right\}(z+h)\right|_{h=0}= \frac 1\eps \langle
W,Q\lambda\rangle+\left\langle
W,\left(\pdr{}{z}+\bk\cdot\nabla_\vx+
\frac{1}{\sqrt{\eps}}
{\cal K}[\hat V,\frac \vx\eps]\right)\lambda\right\rangle,
\end{equation}
hence
\begin{equation}\label{G-eps}
G_\lambda^\eps=\langle W,\lambda(\hat V)\rangle(z)-\int_0^z\left\langle
W,\left(\frac 1\eps Q+ \pdr{}{z}+\bk\cdot\nabla_\vx+
\frac{1}{\sqrt{\eps}}{\cal
K}[\hat V,\frac \vx\eps]\right)\lambda\right\rangle(s)ds
\end{equation}
is a $\tilde P_\eps$-martingale. 
The operator ${\cal K}$ is defined by
\begin{equation}
  \label{eq:Koper}
{\cal K}[\hat V,\veta]\psi(\vx,\veta,\bk,\hat V)=
\frac 1i\int_{\Rm^d} \frac{d\hat V(\bp)}{(2\pi)^d}
e^{i\bp\cdot \veta}
\left[\psi(\vx,\veta,\bk-\frac \bp2)-\psi(\vx,\veta,\bk+\frac \bp2)\right].
\end{equation}
The generator (\ref{generator}) comes from equation (\ref{eq:Weps}) 
written in the form
\begin{equation}
\label{eq:wigner2}
\pdr{W_\eps}{z} + \bk\cdot\nabla_\vx W_\eps=
\frac{1}{\sqrt{\eps}}{\cal K}[\tilde V(z/\eps),\vx/\eps] W_\eps.
\end{equation}

Given a test function $\lambda(z,\vx,\bk)\in C^1([0,L];{\cal S})$ we 
construct a function
\begin{equation}\label{eq:lambdaapprox}
\lambda_\eps(z,\vx,\bk,\hat V)=\lambda(z,\vx,\bk)+
\sqrt{\eps}\lambda_1^\eps(z,\vx,\bk,\hat V)+
\eps\lambda_2^\eps(z,\vx,\bk,\hat V)
\end{equation}
with $\lambda_{1,2}^\eps(t)$ bounded in $L^\infty({\cal
V};{L^2}({\mathbb R}^{2d}))$ uniformly in
$z\in[0,L]$. The functions $\lambda_{1,2}^\eps$ will be chosen so that
\begin{equation}\label{Glambda-Glambdaeps}
\|G_{\lambda_\eps}^\eps(z)-G_\lambda(z)\|_{L^\infty({\cal V})}\le
C_\lambda\sqrt{\eps}
\end{equation}
for all $z\in[0,L]$. Here $G_{\lambda_\eps}^\eps$ is defined by
(\ref{G-eps}) with $\lambda$ replaced by $\lambda_\eps$, and
$G_\lambda$ is defined by (\ref{G-lambda}). The approximate
martingale property (\ref{eq:approxmart}) follows from this.

The functions $\lambda_1^\eps$ and $\lambda_2^\eps$ are as follows.
Let $\lambda_1(z,\vx,\veta,\bk,\hat V)$ be the mean-zero solution of
the Poisson equation
\begin{equation}\label{lambda1eq}
\bk\cdot\nabla_{\veta}\lambda_1+Q\lambda_1=
-{\cal K}\lambda.
\end{equation}
It is given explicitly by
\begin{equation}\label{lambda1-eq}
\lambda_1(z,\vx,\veta,\bk,\hat V)=\frac {1}{i}\int_0^\infty dr e^{rQ}
\int_{\Rm^d} \frac{d\hat V(\bp)}{(2\pi)^{d}}
e^{ir(\bk\cdot \bp)+i(\veta\cdot \bp)}
\left[\lambda(z,\vx,\bk-\frac \bp2)-\lambda(z,\vx,\bk+\frac \bp2)\right].
\end{equation}
Then we let
$\lambda_1^\eps(z,\vx,\bk,\hat V)=\lambda_1(z,\vx,\vx/\eps,\bk,\hat V)$. 
Furthermore, the second order corrector is given by
$\lambda_2^\eps(z,\vx,\bk,\hat V)=\lambda_2(z,\vx,\vx/\eps,\bk,\hat V)$ where
$\lambda_2(z,\vx,\veta,\bk,\hat V)$ is the mean-zero solution of
\begin{equation}\label{lambda2eq}
\bk\cdot\nabla_{\veta}\lambda_2+Q\lambda_2=
{ {\cal L}}\lambda-{\cal K}\lambda_1,
\end{equation}
which exists because $\E\left\{{\cal K}\lambda_1\right\}={ {\cal L}}\lambda$,
and is given by
\begin{eqnarray*}
&&\lambda_2(z,\vx,\veta,\bk,\hat V)=-\int_0^\infty dr e^{rQ}
\left[{\cal L}\lambda(z,\vx,\bk)-
[{\cal K}\lambda_1](z,\vx,\veta+r\bk,\bk,\hat V)\right].
\end{eqnarray*}
Using (\ref{lambda1eq}) and
(\ref{lambda2eq}) we have
\begin{eqnarray*} 
&&\mbox{}\!\!\!\!\!\!\!\!\!\!\!\!
\left.\frac{d}{dh}\E_{W,\hat V,z}^{\tilde P_\eps}
\left\{\langle W,{\lambda_\eps}\rangle\right\}(z+h)\right|_{h=0}
=\left\langle W,\left(\pdr{}{z}+\bk\cdot\nabla_\vx+
\frac{1}{\sqrt{\eps}}{\cal K}[\hat V,\frac \vx\eps]+
\frac 1\eps Q\right)
\left(\lambda+\sqrt{\eps}\lambda_1^\eps+\eps\lambda_2^\eps\right)\right\rangle
\\
&&=\left\langle W, \left(\pdr{}{z}+\bk\cdot\nabla_\vx\right)
  \lambda+{\cal L}\lambda\right\rangle
+\left\langle W, 
   \left(\pdr{}{z}+\bk\cdot\nabla_\vx\right)\left(\sqrt{\eps}\lambda_1^\eps+
\eps\lambda_2^\eps\right)
+
\sqrt{\eps}{\cal K}[\hat V,\frac \vx\eps]\lambda_2^\eps\right\rangle\\
&&=\left\langle W,\left(\pdr{}{z}+\bk\cdot\nabla_\vx\right)\lambda+{\cal
L}\lambda\right\rangle+\sqrt{\eps}\langle W,\zeta_\eps^\lambda\rangle
\end{eqnarray*}
with
\[
\zeta_\eps^\lambda=
\left(\pdr{}{z}+\bk\cdot\nabla_\vx\right)\lambda_1^\eps+
\sqrt{\eps} \left(\pdr{}{z}+\bk\cdot\nabla_\vx\right)\lambda_2^\eps+
{\cal K}[\hat V,\frac \vx\eps]\lambda_2^\eps.
\]
The terms $\bk\cdot\nabla_\vx\lambda_{1,2}^\eps$ above are understood as
differentiation with respect to the slow variable $\vx$ only, and not
with respect to $\veta=\vx/\eps$. It follows that $G_{\lambda_\eps}^\eps$
is given by
\begin{equation}\label{Geps2}
G_{\lambda_\eps}^\eps(z)=\langle W(z),\lambda_\eps\rangle-
\int_0^z ds\left\langle W,\left(\pdr{}{z}+\bk\cdot\nabla_\vx+{\cal L}\right)
\lambda\right\rangle(s)-
\sqrt{\eps}\int_0^z ds\langle W,\zeta_\eps^\lambda\rangle(s)
\end{equation}
and is a martingale with respect to the measure $\tilde P_\eps$
defined on $D([0,L];X\times{\cal V})$, the space of right-continuous
paths with left-side limits \cite{billingsley1}. The estimate
(\ref{eq:approxmart}) follows from the following two lemmas.
\begin{lemma}\label{lemma1new}
Let $\lambda\in C^1([0,L];{\cal S}({\mathbb R}^{2d}))$. Then there
exists a constant $C_\lambda>0$ independent of $z\in[0,L]$ so that the
correctors $\lambda_1^\eps(z)$ and $\lambda_2^\eps(z)$ satisfy the
uniform bounds
\begin{equation}\label{eq:lambda1bdnew}
\|\lambda_1^\eps(z)\|_{{L^\infty({\cal V};L^2)}}+
\|\lambda_2^\eps(z)\|_{{L^\infty({\cal V};L^2)}}\le C_\lambda
\end{equation}
and
\begin{equation}\label{eq:lambda2bdnew}
\Big\|\pdr{\lambda_1^\eps(z)}{z}+
\bk\cdot\nabla_\vx\lambda_1^\eps(z)\Big\|_{{L^\infty({\cal V};L^2)}}+
\Big\|\pdr{\lambda_2^\eps(z)}{z}+
\bk\cdot\nabla_\vx\lambda_2^\eps(z)\Big\|_{{L^\infty({\cal V};L^2)}}\le 
C_\lambda.
\end{equation}
\end{lemma}
\begin{lemma}\label{lemma3}
There exists a constant $C_\lambda$ such that
\[
\|{\cal K}[\hat V,\vx/\eps]\|_{L^2\to L^2} \le C
\]
for any $\hat V\in{\cal V}$ and all $\eps\in(0,1]$.
\end{lemma}

Indeed, (\ref{eq:lambda1bdnew}) implies that $\left|\langle W,\lambda\rangle
- \langle W,\lambda_\eps\rangle\right|\le C\sqrt{\eps}$ for
all $W\in X$ and $\hat V\in{\cal V}$, while (\ref{eq:lambda2bdnew}) and
Lemma \ref{lemma3} imply that for all $z\in[0,L]$
\begin{equation}\label{zetabd}
\|\zeta_\eps^\lambda(z)\|_{L^2}\le C
\end{equation}
for all $\hat V\in{\cal V}$ so that (\ref{eq:approxmart}) follows.

{\bf Proof of Lemma \ref{lemma3}.}  Lemma \ref{lemma3} follows
immediately from the definition of ${\cal K}$, the bound (\ref{defcalv})
and the Cauchy-Schwartz inequality.

We now prove Lemma \ref{lemma1new}. We will omit the $z$-dependence
of the test function $\lambda$ to simplify the
notation.

{\bf Proof of Lemma \ref{lemma1new}.} We only prove
(\ref{eq:lambda1bdnew}). Since $\lambda\in{\cal S}({\mathbb R}^{2d})$,
there exists a constant $C_\lambda$ so that
\[
|\lambda(\vx,\bk)|\le\frac{C_\lambda}{(1+|\vx|^{5d})(1+|\bk|^{5d})}.
\]
Then we obtain using (\ref{defcalv}) and (\ref{eq:expdecaynew})
\begin{eqnarray*}
&&|\lambda_1^\eps(z,\vx,\bk,\hat V)|=C\left|\int_0^\infty dr e^{rQ}
\int_{\Rm^d} d\hat V(\bp)e^{ir(\bk\cdot\bp)+i(\vx\cdot\bp)/\eps}\left[
\lambda(z,\vx,\bk-\frac{\bp}{2})-\lambda(z,\vx,\bk+\frac{\bp}{2})\right]\right|
\\
&&\le {C}\int_0^\infty dr e^{-\alpha r}\sup_{\hat V}
  \int_{\Rm^d} |d\hat V(\bp)|
\left[|\lambda(z,\vx,\bk-\frac{\bp}{2})|+
|\lambda(z,\vx,\bk+\frac{\bp}{2})|\right]\\
&&\le
\frac{C}{(1+|\vx|^{5d})(1+(|\bk|-L)^{5d}\chi_{|\bk|\ge 5L}(\bk))}
\end{eqnarray*}
and the $L^2$-bound on $\lambda_1$ follows.

We show next that $\lambda_2^\eps$ is uniformly bounded. We have
\begin{eqnarray*}
&&\lambda_2^\eps(\vx,\bk,\hat V)=-\int_0^\infty dr e^{rQ}
\left[{\cal L}\lambda(\vx,\bk)-
\frac 1i\int_{\Rm^d} \frac{d\hat V(\bp)}{(2\pi)^d}
 e^{i\bp\cdot(\vx/\eps+r\bk)}\right.\\
&&\left.~~~~~~~~~~~~~~~
\times\left[\lambda_1(\vx,\frac \vx\eps+r\bk,\bk-\frac \bp2,\hat V)-
\lambda_1(\vx,\frac {\vx}{\eps}+r\bk,\bk+\frac \bp2,\hat V)\right]\right].
\end{eqnarray*}
The second term above may be written as
\begin{eqnarray*}
&&\frac 1i\int_{\Rm^d}\frac{d\hat V(\bp)}{(2\pi)^d}
e^{i\bp\cdot(\vx/\eps+r\bk)}
\left[\lambda_1(\vx,\frac \vx\eps+r\bk,\bk-\frac \bp2,\hat V)-
\lambda_1(\vx,\frac {\vx}{\eps}+r\bk,\bk+\frac \bp2,\hat V)\right]\\
&&=-\int_{\Rm^d} \frac{d\hat V(\bp)}{(2\pi)^{d}}e^{i\bp\cdot(\bx/\eps+r\bk)}
\int_0^\infty ds e^{sQ}\int_{\Rm^d} \frac{d\hat V(\bq)}{(2\pi)^d}
e^{is(\bk-\bp/2)\cdot \bq+i(\bx/\eps+r\bk)\cdot \bq}\\
&&\times
\left[\lambda(\vx,\bk-\frac \bp2-\frac \bq2)-
\lambda(\vx,\bk-\frac \bp2+\frac \bq2)\right]
\\
&&+
\int_{\Rm^d} \frac{d\hat V(\bp)}{(2\pi)^d}
e^{i\bp\cdot(\bx/\eps+r\bk)}
\int_0^\infty ds e^{sQ}\int_{\Rm^d} \frac{d\hat V(\bq)}{(2\pi)^d}
e^{is(\bk+\bp/2)\cdot \bq+i(\bx/\eps+r\bk)\cdot \bq}\\
&&\times\left[
\lambda(\vx,\bk+\frac \bp2-\frac \bq2)-
\lambda(\vx,\bk+\frac \bp2+\frac \bq2)\right].
\end{eqnarray*}
Therefore we obtain
\begin{eqnarray*}
&&|\lambda_2^\eps(\vx,\bk,\hat V)|\le C\int_0^\infty{dr}e^{-\alpha r}
\left[|{\cal L}\lambda(\vx,\bk)|+\sup_{\hat V}\int_{\Rm^d}{|d\hat V(\bp)|}
\int_0^\infty ds e^{-\alpha s}\sup_{\hat V_1}
  \int_{\Rm^d} |d\hat V_1(\bq)|\right.\\
&&\left.\times\left(|\lambda(\vx,\bk-\frac \bp2-\frac \bq2)|+
|\lambda(\vx,\bk-\frac \bp2+\frac \bq2)|+
|\lambda(\vx,\bk+\frac \bp2-\frac \bq2)|+
\lambda(\vx,\bk+\frac \bp2+\frac \bq2)\right)\right]\\
&&\le C\left[|{\cal L}\lambda(\vx,\bk)|+
\frac{1}{(1+|\vx|^{5d})(1+(|\bk|-L)^{5d}\chi_{|\bk|\ge 5L}(\bk))}\right]
\end{eqnarray*}
and the $L^2$-bound on $\lambda_2^\eps$ in (\ref{eq:lambda1bdnew}) follows
because the operator ${\cal L}:L^2\to L^2$ is bounded. The proof of
(\ref{eq:lambda2bdnew}) is very similar and is omitted.

%The process $W_\eps(z)$ generates a
%probability measure $P_\eps$ on the space $C([0,T];X)$ with 
%$X=\left\{W\in{L^2}:~\|W\|_{L^2}\le C\right\}$. This family is tight.
%\begin{lemma}\label{lemma-tight}
%The family of measures $P_\eps$ is weakly compact.
%\end{lemma}
%The proof of this lemma may be found in \cite{BPR-Nonlin}.

Lemma \ref{lemma1new} and Lemma \ref{lemma3} together with
(\ref{Geps2}) imply the bound (\ref{Glambda-Glambdaeps}). The
tightness of measures $P_\eps$ given by Lemma \ref{lemma-tight}
implies then that the expectation $\E\left\{W_\eps(z,\vx,\bk)\right\}$
converges weakly in $L^2({\mathbb R}^{2d})$ to the solution $\overline
W(z,\vx,\bk)$ of the transport equation for each $z\in[0,L]$.

\subsection{Convergence in probability}
\label{sec:cvproba}

We now prove that for any test function $\lambda$ the second moment
$\E\left\{\langle W_\eps,\lambda\rangle^2\right\}$ converges to
$\langle \overline W,\lambda\rangle^2$. This will imply the convergence in
probability claimed in Theorem \ref{thm2}. The proof is similar to
that for $\E\left\{\langle W_\eps,\lambda\rangle\right\}$ and is based
on constructing an appropriate approximate martingale for the
functional $\langle W\otimes W,\mu\rangle$, where
$\mu(z,\vx_1,\bk_1,\vx_2,\bk_2)$ is a test function, and $W\otimes
W(z,\vx_1,\bk_1,\vx_2,\bk_2)=W(z,\vx_1,\bk_1)W(z,\vx_2,\bk_2)$.  We
need to consider the action of the infinitesimal generator on
functions of $W$ and $\hat V$ of the form
\begin{displaymath}
F(W,\hat V)=\langle 
W(\vx_1,\bk_1)W(\vx_2,\bk_2),\mu(z,\vx_1,\bk_1,\vx_2,\bk_2,\hat V)\rangle=
\langle W\otimes W,\mu(\hat V)\rangle
\end{displaymath}
where $\mu$ is a given function.
The infinitesimal generator acts on such functions as
\begin{eqnarray}\label{generator2}
&&\left.\frac{d}{dh}\E_{W,\hat V,z}^{\tilde P_\eps}\left\{\langle
W\otimes W,\mu(\hat V)\rangle\right\}(z+h)\right|_{h=0}= 
\dfrac1\eps\langle W\otimes W,Q\lambda\rangle+
\langle W\otimes W, {\cal H}_{2}^\eps\mu\rangle,
\end{eqnarray}
where
\begin{equation}
  \label{eq:Heps}
{\cal H}_{2}^{\eps}\mu=\dsum_{j=1}^2 \dfrac{1}{\sqrt\eps}{\cal K}_j
\left[\hat V,\dfrac{\vx^j}{\eps}\right]\mu+ \bk^j\cdot\nabla_{\vx^j}\mu,
\end{equation}
with
\[
{\cal K}_1[\hat V,\veta_1]\mu=\frac 1i
 \int_{\Rm^d} d\hat V(\bp)e^{i(\bp\cdot \veta_1)}
\left[\mu(\bk_1-\frac \bp2,\bk_2)-\mu(\bk_1+\frac \bp2,\bk_2)\right]
\]
and
\[
{\cal K}_2[\hat V,\veta_2]\mu=\frac 1i
 \int_{\Rm^d} d\hat V(\bp)e^{i(\bp\cdot \veta_2)}
\left[\mu(\bk_1,\bk_2-\frac \bp2)-\mu(\bk_1,\bk_2+\frac \bp2)\right].
\]
% \\
%&&~~~~~~~~+ \langle W\otimes W,
%\Big(\pdr{}{z}+\bk_1\cdot\nabla_{\vx_1} + \bk_2\cdot\nabla_{\vx_2}+
%\dfrac{1}{\sqrt{\eps}}({\cal K}_1[\hat V,\dfrac{\vx_1}{\eps}] 
%  +{\cal K}_2[\hat V,\dfrac{\vx_2}{\eps}] )\Big) \mu\rangle.\nonumber
%\end{eqnarray}
Therefore the functional
\begin{eqnarray}\label{G-eps2}
&&G_\mu^{2,\eps}=\langle W\otimes W,\mu(\hat V)\rangle(z)\\
&&-
\int_0^z\left\langle W\otimes W,
\Big(\frac{1}{\eps}Q+
\pdr{}{z}+\bk_1\cdot\nabla_{\vx_1} + \bk_2\cdot\nabla_{\vx_2}+
\dfrac{1}{\sqrt{\eps}}({\cal K}_1[\hat V,\dfrac{\vx_1}{\eps}] 
  +{\cal K}_2[\hat V,\dfrac{\vx_2}{\eps}] )\Big) \mu\right\rangle(s)ds 
\nonumber
\end{eqnarray}
is a $\tilde P^\eps$ martingale. We let $\mu(z,\vX,\bK)\in{\cal
S}({\mathbb R}^{2d}\times{\mathbb R}^{2d})$ be a test function
independent of $\hat V$, where $\vX=(\vx_1,\vx_2)$, and
$\bK=(\bk_1,\bk_2)$. We define an approximation
\begin{displaymath}
  \mu_\eps(z,\vX,\bK)=\mu(z,\vX,\bK)+
\sqrt{\eps}\mu_1(z,\vX,\vX/\eps,\bK)+\eps\mu_2(\vX,\vX/\eps,\bK).
\end{displaymath}
We will use the notation
$\mu_1^\eps(z,\vX,\bK)=\mu_1(z,\vX,\vX/\eps,\bK)$ and
$\mu_2^\eps(z,\vX,\bK)=\mu_2(z,\vX,\vX/\eps,\bK)$.
% and $W_{2,\mu}^\eps=\langle W\otimes W,\mu_\eps\rangle$. 
The functions $\mu_1$ and $\mu_2$ are to be determined.  We now use
(\ref{generator2}) to get
%\begin{displaymath}
%F_\eps(W,V)=\langle W_2,\lambda_0+\sqrt\eps\lambda_1^{\eps}
%+\eps \lambda_2^{\eps}\rangle,
%\end{displaymath}
%Since $F_\eps$ is a sum of functions of
%the form (\ref{eq:formF}), we obtain that
\begin{eqnarray}
\label{eq:derivFeps}
&&D_\eps:=\dfrac{d}{dh}\Big|_{h=0}\E_{W,\hat V,z}(\langle W\otimes
W,\mu_\eps(\hat V))(z+h) =
\dfrac{1}{\eps} \left\langle W\otimes W,
\left(Q+\dsum_{j=1}^2\bk^j\cdot\nabla_{\veta^j}\right)\mu\right\rangle \\
&&+ \dfrac{1}{\sqrt\eps} \left\langle W\otimes W,
\left(Q+\dsum_{j=1}^2  \bk^j\cdot\nabla_{\veta^j} \right)
\mu_1+\dsum_{j=1}^2 {\cal K}_j\left[\hat V,\veta^j\right]\mu
\right\rangle \nonumber\\
&&+  \left\langle  W\otimes W,
\left(Q+\dsum_{j=1}^2  \bk^j\cdot\nabla_{\veta^j} \right) \mu_2
+ \dsum_{j=1}^2 {\cal K}_j\left[\hat V,\veta^j\right] \mu_1
+ \pdr{\mu}{z}+
\dsum_{j=1}^2\bk^j\cdot\nabla_{\vx^j} \mu  \right\rangle \nonumber\\
&&+ \sqrt\eps  \left\langle W\otimes W,
\dsum_{j=1}^2 {\cal K}_j\left[\hat V,\veta^j\right] \mu_2
+ \left(\pdr{}{z}+\dsum_{j=1}^2\bk^j\cdot\nabla_{\vx^j}\right)
(\mu_1+\sqrt\eps\mu_2)\right\rangle\nonumber. 
\end{eqnarray}
The above expression is evaluated at $\veta_j=\vx_j/\eps$.  The term
of order $\eps^{-1}$ in $D_\eps$ vanishes since $\mu$ is independent
of $V$ and the fast variable $\veta$. We cancel the term of order
$\eps^{-1/2}$ in the same way as before by defining $\mu_1$ as the
unique mean-zero (in the variables $\hat V$ and
$\veta=(\veta_1,\veta_2)$) solution of
\begin{equation}\label{eq:lambda1} 
\big(Q+\dsum_{j=1}^2 \bk^j\cdot\nabla_{\veta^j} \big)
\mu_1 + \dsum_{j=1}^2 {\cal K}_j\Big[\hat V,\veta^j\Big]\mu=0.
\end{equation}
It is given explicitly by
\begin{eqnarray*}
&&\mu_1(\vX,\veta,\bK,\hat V)=\frac 1i\int_0^\infty dr e^{rQ}
  \int_{\Rm^d} d\hat V(\bp)
e^{ir(\bk_1\cdot \bp)+i(\veta_1\cdot \bp)}\left[\mu(\bk_1-\frac \bp2,\bk_2)-
\mu(\bk_1+\frac \bp2,\bk_2)\right]\\
&&~~~~~~~~~~~~~~~~~~+
\frac 1i\int_0^\infty dr e^{rQ}\int_{\Rm^d} d\hat V(\bp)
e^{ir(\bk_2\cdot \bp)+i(\veta_2\cdot \bp)}\left[\mu(\bk_1,\bk_2-\frac \bp2)-
\mu(\bk_1,\bk_2+\frac \bp2)\right].
\end{eqnarray*}
When $\mu$ has the form $\mu=\lambda\otimes\lambda$, then $\mu_1$ has the form
$\mu_1=\lambda_1\otimes\lambda+\lambda\otimes\lambda_1$
with the corrector $\lambda_1$ given by (\ref{lambda1-eq}).
Let us also define $\mu_2$ as the mean zero
with respect to $\pi_V$ solution of
\begin{equation}\label{eq:lambda2}
\big(Q+\dsum_{j=1}^2  \bk^j\cdot\nabla_{\veta^j} \big) \mu_2
     + \dsum_{j=1}^2 {\cal K}_j\Big[\hat V,\veta^j\Big] \mu_1
        = \overline{\dsum_{j=1}^2 {\cal K}_j\Big[\hat V,\veta^j\Big] \mu_1},
\end{equation}
where $\overline f=\int d\pi_Vf$. The function $\mu_2$ is given by
\begin{eqnarray}\label{lambda2long}
&&\mu_2(\vX,\veta,\bK,\hat V)=-\int_0^\infty dr e^{rQ}
\big[\overline{{\cal K}_1[\hat V,\veta_1+r\bk_1]
\mu_1(\vX,\veta+r\bK,\bK)}\\
&&~~~~~~~~~~~~~~~~~~~~-[{\cal K}_1[\hat V,\veta_1+r\bk_1]
\mu_1](\vX,\veta+r\bK,\bK,\hat V)\big]\nonumber\\
&&~~~~~~~~~~~~~~~~~~~~-
\int_0^\infty dr e^{rQ}
\big[\overline{{\cal K}_2[\hat V,\bk_2+r\veta_2]
\mu_1(\vX,\veta+r\bK,\bK)}
\nonumber
\\ &&~~~~~~~~~~~~~~~~~~~~
-[{\cal K}_2[\hat V,\veta_2+r\bk_2]\mu_1]
  (\vX,\veta+r\bK,\bK,\hat V)
\big].\nonumber
\end{eqnarray}
Unlike the first corrector $\mu_1$, the second corrector $\mu_2$ may
not be written as an explicit sum of tensor products even if $\mu$ has
the form $\mu=\lambda\otimes\lambda$ because $\mu_1$ depends on $\hat V$. 

The $\tilde P^\eps$-martingale $G_{\mu_\eps}^{2,\eps}$ is given by
\begin{eqnarray}\label{G-eps2-2}
&&G_\mu^{2,\eps}=\langle W\otimes W,\mu(\hat V)\rangle(z)-
\int_0^z\left\langle W\otimes W,
\Big(
\pdr{}{z}+\bk_1\cdot\nabla_{\vx_1} + \bk_2\cdot\nabla_{\vx_2}+
{\cal L}_2^\eps\Big) \mu\right\rangle(s)ds\\
&&-\sqrt\eps
\dint_0^z  \langle W\otimes W,\zeta^\mu_\eps\rangle(s) ds,
\nonumber
\end{eqnarray}
where $\zeta^\mu_\eps$ is given by
\begin{eqnarray*}
\zeta_\mu^\eps=\dsum_{j=1}^2 
{\cal K}_j\left[\hat V,\frac{\vx_j}{\eps}\right]\mu_2^\eps
+ \left(\pdr{}{z}+\dsum_{j=1}^2\bk^j\cdot\nabla_{\vx^j}\right)
(\mu_1^\eps+\sqrt\eps\mu_2^\eps)
\end{eqnarray*}
and the operator ${\cal L}_2^\eps$ is defined by
\begin{equation} \label{eq:L-2eps}
  \begin{array}{l}
{\cal L}_2^\eps\mu=-\dfrac{1}{(2\pi)^{d}}
 \dint_0^\infty dr \dint_{\Rm^d} d\bp \tilde R(r,\bp)
  \Big[e^{ir(\bk_1+\frac{\bp}{2})\cdot \bp} (\mu(z,\bx_1,\bk_1,\bx_2,\bk_2)-
   \mu(z,\bx_1,\bk_1+\bp,\bx_2,\bk_2) ) \\-
  e^{ir(\bk_1-\frac{\bp}{2})\cdot \bp} (\mu(z,\bx_1,\bk_1-\bp,\bx_2,\bk_2)-
   \mu(z,\bx_1,\bk_1,\bx_2,\bk_2) ) \Big] \\
 + \Big[e^{i\bp\cdot\frac{\bx_2-\bx_1}{\eps}+ir\bk_2\cdot \bp}
  (\mu(z,\bx_1,\bk_1+\frac{\bp}{2},\bx_2,\bk_2-\frac{\bp}{2})-
   \mu(z,\bx_1,\bk_1+\frac{\bp}{2},\bx_2,\bk_2+\frac{\bp}{2}))\\
       -e^{i\bp\cdot \frac{\bx_2-\bx_1}{\eps}+ir\bk_2\cdot \bp}
  (\mu(z,\bx_1,\bk_1-\frac{\bp}{2},\bx_2,\bk_2-\frac{\bp}{2})-
   \mu(z,\bx_1,\bk_1-\frac{\bp}{2},\bx_2,\bk_2+\frac{\bp}{2}))\Big]\\
 + \Big[e^{ir\bk_1\cdot \bp + i\frac{\bx_1-\bx_2}{\eps}\cdot \bp}
  (\mu(z,\bx_1,\bk_1-\frac{\bp}{2},\bx_2,\bk_2+\frac{\bp}{2})-
   \mu(z,\bx_1,\bk_1-\frac{\bp}{2},\bx_2,\bk_2-\frac{\bp}{2}))\\
       -e^{ir\bk_1\cdot \bp + i\frac{\bx_1-\bx_2}{\eps}\cdot \bp}
  (\mu(z,\bx_1,\bk_1+\frac{\bp}{2},\bx_2,\bk_2+\frac{\bp}{2})-
   \mu(z,\bx_1,\bk_1+\frac{\bp}{2},\bx_2,\bk_2-\frac{\bp}{2}))\Big] \\
  +\Big[e^{ir(\bk_2+\frac{\bp}{2})\cdot \bp}
  (\mu(z,\bx_1,\bk_1,\bx_2,\bk_2)-\mu(z,\bx_1,\bk_1,\bx_2,\bk_2+\bp))\\
  -e^{ir(\bk_2-\frac{\bp}{2})\cdot \bp}
  (\mu(z,\bx_1,\bk_1,\bx_2,\bk_2-\bp)-\mu(z,\bx_1,\bk_1,\bx_2,\bk_2))\Big].
  \end{array}
\end{equation}
We have used in the calculation of ${\cal L}_2^\eps$ that for a
sufficiently regular function $f$, we have
\begin{displaymath}
  \E\left[\dint_{\Rm^d} \frac{d\hat V(\bq)}{(2\pi)^{d}}
   \dint_0^\infty  dr\,e^{rQ} \dint_{\Rm^d} {d\hat V(\bp)}
     f(r,\bp,\bq)\right]
   =\dint_0^\infty dr \dint_{\Rm^d} 
      \tilde R(r,\bp) f(r,\bp,-\bp) d\bp.
\end{displaymath}
The bound on $\zeta^\mu_\eps$ is similar to that on
$\zeta^\lambda_\eps$ obtained previously as the correctors $\mu_j^\eps$
satisfy the same kind of estimates as the correctors $\lambda_j$:  
\begin{lemma}\label{lemma8}
There exists a constant $C_\mu>0$ so that the functions
$\mu_{1,2}^\eps$ obey the uniform bounds
\begin{equation}\label{eq:lambda1M}
\|\mu_1^\eps(z)\|_{L^2({\mathbb R}^{2d})}+
\|\mu_2^\eps\|_{L^2({\mathbb R}^{2d})}\leq C_\mu
\end{equation}
and
\begin{equation}\label{eq:lambda4bdnew}
\Big\|\pdr{\mu_1^\eps(z)}{z}+
\dsum_{j=1}^2\bk_j\cdot\nabla_{\vx_j}
\mu_1^\eps(z)\Big\|_{{L^2({\mathbb R}^{2d})}}+
\Big\|\pdr{\mu_2^\eps(z)}{z}+
\dsum_{j=1}^2\bk_j\cdot\nabla_{\vx_j}
\mu_2^\eps(z)\Big\|_{{L^2({\mathbb R}^{2d})}}\le 
C_\mu
\end{equation}
for all $z\in[0,L]$ and $V\in{\cal V}$.
\end{lemma}
The proof of this lemma is very similar to that of Lemma \ref{lemma1new}
and is therefore omitted.

Unlike the first moment case, the averaged operator ${\cal L}_2^\eps$
still depends on $\eps$. We therefore do not have strong convergence
of the $\tilde P^\eps$-martingale $G_{\mu_\eps}^{2,\eps}$ to its limit
yet. However, the a priori bound on $W_\eps$ in $L^2$ allows us to
characterize the limit of $G_{\mu_\eps}^{2,\eps}$ and show strong
convergence. This is shown as follows.  The first and last terms in
(\ref{eq:L-2eps}) that are independent of $\eps$ give the contribution:
\begin{eqnarray*}
&&  {\cal L}_2\mu=
 \dint_0^\infty dr \dint_{\Rm^d}
   \dfrac{d\bp}{(2\pi)^{d}}\left[\tilde R(r,\bp-\bk_1)
e^{ir\frac{p^2-k_1^2}{2}}
(\mu(z,\bx_1,\bp,\bx_2,\bk_2)-\mu(z,\bx_1,\bk_1,\bx_2,\bk_2))\right.\\
&&+ \tilde R(r,\bk_1-\bp)e^{ir\frac{k_1^2-p^2}{2}}
 (\mu(z,\bx_1,\bp_1,\bx_2,\bk_2)-\mu(z,\bx_1,\bk_1,\bx_2,\bk_2))\\
&&+\tilde R(z,\bp-\bk_2)e^{ir\frac{p^2-k_2^2}{2}}
(\mu(z,\bx_1,\bk_1,\bx_2,\bp)-\mu(z,\bx_1,\bk_1,\bx_2,\bk_2))\\
&&\left.+\tilde R(z,\bk_2-\bp)e^{ir\frac{k_2^2-p^2}{2}}
(\mu(z,\bx_1,\bk_1,\bx_2,\bp)-\mu(z,\bx_1,\bk_1,\bx_2,\bk_2))\right]\\
&&  =\dint_{\Rm^d}
   \dfrac{d\bp}{(2\pi)^{d}}  \hat R(\frac{p^2-k_1^2}{2},\bp-\bk_1)
(\mu(z,\vx_1,\bp,\vx_2,\bk_2)-\mu(z,\vx_1,\bk_1,\vx_2,\bk_2))\\
&& + \hat R(\dfrac{p^2-k_2^2}{2},\bp-\bk_2)
(\mu(z,\vx_1,\bk_1,\vx_2,\bp)-\mu(z,\vx_1,\bk_1,\vx_2,\bk_2)). 
\end{eqnarray*}
The two remaining terms give a contribution that tends to $0$ as
$\eps\to0$ for sufficiently smooth test functions.  They are given by
\begin{displaymath}
  \begin{array}{l}
({\cal L}_2^\eps-{\cal L}_2)\mu= \dfrac{1}{(2\pi)^{d}}
 \dint_0^\infty dr \dint_{\Rm^d} d\bp \tilde R(r,\bp) \times\\
\Big(e^{i\bp\cdot\frac{\bx_2-\bx_1}{\eps}+ir\bk_2\cdot \bp}
+e^{ir\bk_1\cdot \bp + i\frac{\bx_1-\bx_2}{\eps}\cdot \bp}\Big)
\Big(\mu(z,\bx_1,\bk_1+\frac{\bp}{2},\bx_2,\bk_2+\frac{\bp}{2})-
   \mu(z,\bx_1,\bk_1+\frac{\bp}{2},\bx_2,\bk_2-\frac{\bp}{2})\Big)\\
+\Big(e^{i\bp\cdot\frac{\bx_2-\bx_1}{\eps}+ir\bk_2\cdot \bp}
+ e^{ir\bk_1\cdot \bp + i\frac{\bx_1-\bx_2}{\eps}\cdot \bp}\Big)
\Big(\mu(z,\bx_1,\bk_1-\frac{\bp}{2},\bx_2,\bk_2-\frac{\bp}{2})\!-
   \mu(z,\bx_1,\bk_1-\frac{\bp}{2},\bx_2,\bk_2+\frac{\bp}{2})\Big).
  \end{array}
\end{displaymath}
We have
%\begin{displaymath}
%  R(z,\bx)=R(-z,\bx)=R(z,-\bx)\in\Rm
%\end{displaymath}
%This implies that
\begin{displaymath}
  \tilde R(z,\bp)=\tilde R(-z,-\bp)\ge 0
\end{displaymath}
by Bochner's theorem.
Since $({\cal L}_2^\eps-{\cal L}_2)$ and $\lambda$ are real
quantities, we can take the real part of the above term and, after the
change of variables $r\to-r$ and $\bp\to-\bp$, obtain
\begin{displaymath}
  \begin{array}{l}
({\cal L}_2^\eps-{\cal L}_2)\mu= \dfrac{1}{(2\pi)^{d}}
 \dint_{-\infty}^\infty dr \dint_{\Rm^d} d\bp \tilde R(r,\bp)
\cos(\bp\cdot\frac{\bx_2-\bx_1}{\eps})
(e^{ir\bk_2\cdot \bp}+e^{ir\bk_1\cdot \bp})\\
~~~~~~~~~~~~~~\times
\Big(\mu(z,\bx_1,\bk_1+\frac{\bp}{2},\bx_2,\bk_2+\frac{\bp}{2})+
 \mu(z,\bx_1,\bk_1-\frac{\bp}{2},\bx_2,\bk_2-\frac{\bp}{2})\\
~~~~~~~~~~~~~~-\mu(z,\bx_1,\bk_1+\frac{\bp}{2},\bx_2,\bk_2-\frac{\bp}{2})-
 \mu(z,\bx_1,\bk_1-\frac{\bp}{2},\bx_2,\bk_2+\frac{\bp}{2})\Big)\\
~~~~~~~~~~~~~~=\dfrac{2}{(2\pi)^{d}}
 {\dint_{\Rm^d}} d\bp (\hat R(-\bk_1\cdot \bp,\bp)+\hat R(-\bk_2\cdot \bp,\bp))
   \cos(\bp\cdot\frac{\bx_2-\bx_1}{\eps})\\
~~~~~~~~~~~~~~
\times\Big(\mu(z,\bx_1,\bk_1+\frac{\bp}{2},\bx_2,\bk_2+\frac{\bp}{2})-
 \mu(z,\bx_1,\bk_1-\frac{\bp}{2},\bx_2,\bk_2+\frac{\bp}{2})\Big)\\
~~~~~~~~~~~~~~=g_1+g_2+g_3+g_4 +c.c.
  \end{array}
\end{displaymath}
We have (since $\mu$ is real-valued)
\begin{eqnarray*}
\!\!\!\!\!\!&&I=\!\int_{\Rm^{4d}} \!\!\!\!\!\!d\vx_1d\bk_1d\vx_2d\bk_2
  |g_1(z,\vx_1,\bk_1,\vx_2,\bk_2)|^2=
C\int_{\Rm^{6d}}\!\!\!\!\!\! d\vx_1d\bk_1d\vx_2d\bk_2d\bp d\bq
\hat R(-\bk_1\cdot \bp,\bp)\hat R(-\bk_1\cdot \bq,\bq)\\
&&\times
e^{i(\bp-\bq)\cdot\frac{\bx_2-\bx_1}{\eps}}
 \mu(z,\bx_1,\bk_1-\frac{\bp}{2},\bx_2,\bk_2+\frac{\bp}{2})
\mu(z,\bx_1,\bk_1-\frac{\bq}{2},\bx_2,\bk_2+\frac{\bq}{2}).
\end{eqnarray*}
Using density arguments we may assume that $\mu$ has the form
\[
\mu(\bx_1,\bk_1,\bx_2,\bk_2)=\mu_1(\bx_1-\bx_2)\mu_2(\bx_1+\bx_2)
\mu_3(\bk_1)\mu_4(\bk_2).
\]
%Let $\chi(r)$ be a smooth function so that $0\le\chi(r)\le 1$,
%$\chi(r)=0$ if $|r|\le 1$ and $\chi(r)=1$ if $|r|\ge 2$. Then we have
%$I=I_1+I_2$ where
%\begin{eqnarray*}
%&&I_1=C\int d\vx_1d\bk_1d\vx_2d\bk_2d\bp d\bq
%\hat R(-\bk_1\cdot \bp,\bp)\hat R(-\bk_1\cdot \bq,\bq)
%\left[1-\chi\left(\frac{|\bp-\bq|}{\delta}\right)\right]\\
%&&\times
%e^{i(\bp-\bq)\cdot\frac{\bx_2-\bx_1}{\eps}}
% \mu(z,\bx_1,\bk_1-\frac{\bp}{2},\bx_2,\bk_2+\frac{\bp}{2})
%\mu(z,\bx_1,\bk_1-\frac{\bq}{2},\bx_2,\bk_2+\frac{\bq}{2})
%\end{eqnarray*}
%so that
%\begin{eqnarray*}
%&&|I_1|\le C\|\mu(z)\|_{L^2({\mathbb R}^{2d})}\int d\bp d\bq G(\bp)G(\bq)
%\left[1-\chi\left(\frac{|\bp-\bq|}{\delta}\right)\right]\to 0
%\end{eqnarray*}
%as $\delta\to 0$, with $G(\bp)=\sup_\omega \hat R(\omega,\bp)$.
Then we have
\begin{eqnarray*}
&&I=C\int_{\Rm^{6d}}\!\!\!\!\! d\vx_1d\bk_1d\vx_2d\bk_2d\bp d\bq
\hat R(-\bk_1\cdot \bp,\bp)\hat R(-\bk_1\cdot \bq,\bq)\\
&&\times
e^{-i(\bp-\bq)\cdot\frac{\bx_1}{\eps}}
 \mu_1^2(\vx_1)\mu_2^2(\vx_2)
\mu_3(\bk_1-\frac{\bp}{2})\mu_4(\bk_2+\frac{\bp}{2})
\mu_3(\bk_1-\frac{\bq}{2})\mu_4(\bk_2+\frac{\bq}{2})\\
&&=C\|\mu_2\|_{L^2}^2\int_{\Rm^{4d}}\!\!\! d\bk_1d\bk_2d\bp d\bq
\hat R(-\bk_1\cdot \bp,\bp)\hat R(-\bk_1\cdot \bq,\bq)
\hat\nu(\frac{\bp-\bq}{\eps})\\
&&\times\mu_3(\bk_1-\frac{\bp}{2})\mu_4(\bk_2+\frac{\bp}{2})
\mu_3(\bk_1-\frac{\bq}{2})\mu_4(\bk_2+\frac{\bq}{2})
\end{eqnarray*}
where $\nu(\vx)=\mu_1^2(\vx)$. We introduce $G(\bp)=\sup_\omega \hat
R(\omega,\bp)$ and use the Cauchy-Schwartz inequality in $\bk_1$ and $\bk_2$:
\begin{eqnarray*}
&&|I|\le C\|\mu_2\|_{L^2}^2\|\mu_3\|_{L^2}^2\|\mu_4\|_{L^2}^2
\int_{\Rm^{2d}}\! d\bp d\bq G(\bp)G(\bq)
\left|\hat\nu(\frac{\bp-\bq}{\eps})\right|.
\end{eqnarray*}
We use again the Cauchy-Schwartz inequality, now in $\bp$, to get
\begin{eqnarray*}
&&|I|\le C\|\mu_2\|_{L^2}^2\|\mu_3\|_{L^2}^2\|\mu_4\|_{L^2}^2\|G\|_{L^2}
\int_{\Rm^d}\! d\bq G(\bq)
\left(\int_{\Rm^d} d\bp\left|\hat\nu(\frac{\bp}{\eps})\right|^2\right)^{1/2}\\
&&\le C\eps^{d/2}
\|\mu_2\|_{L^2}^2\|\mu_3\|_{L^2}^2\|\mu_4\|_{L^2}^2\|G\|_{L^2}\|G\|_{L^1}
\|\nu\|_{L^2}.
\end{eqnarray*}
This proves that $\|({\cal L}_2^\eps-{\cal L}_2)\mu\|_{L^2}\to 0$ as
$\eps\to 0$. Notice that oscillatory integrals of the form
\[
\int_{{\Rm}^d} e^{i\frac{\bp\cdot\vx}{\eps}}\mu(\bp)d\bp
\]
are not small in the bigger space ${\cal A}'$, which is natural in the
context of Wigner transforms and was used in \cite{BPR-Nonlin}. In
this bigger space, we cannot control $({\cal L}_2^\eps-{\cal L}_2)\mu$
and actually suspect that the limit measure $P$ may no longer be
deterministic.
%The space ${\cal A}'$ is the natural space for the Wigner distribution
%of solutions of the Schr\"odinger   

We therefore deduce that
\[
G_\mu^{2}=\langle W\otimes W,\mu(\hat V)\rangle(z)-
\int_0^z\left\langle W\otimes W,
\Big(
\pdr{}{z}+\bk_1\cdot\nabla_{\vx_1} + \bk_2\cdot\nabla_{\vx_2}+
{\cal L}_2\Big) \mu\right\rangle(s)ds
\] 
is an approximate $\tilde P_\eps$ martingale.  The limit of the second
moment 
\begin{displaymath}
  W_2(z,\bx_1,\bk_1,\bx_2,\bk_2)=
\E^P\left\{W(z,\bx_1,\bk_1)W(z,\bx_2,\bk_2)\right\}
\end{displaymath}
thus satisfies (weakly) the transport equation
\begin{displaymath}
  \pdr{W_2}{t}+(\bk_1\cdot\nabla_{\bx_1}+\bk_2\cdot\nabla_{\bx_2}) W_2
 = {\cal L}_2 W_2,
\end{displaymath}
with initial data
$W_2(0,\bx_1,\bk_1,\bx_2,\bk_2)=W_0(\bx_1,\bk_1)W_0(\bx_2,\bk_2)$.
Moreover, the operator ${\cal L}_2$ acting on a tensor product
$\lambda\otimes\lambda$ has the form
\[
{\cal L}_2[\lambda\otimes\lambda]={\cal L}\lambda\otimes\lambda+
\lambda\otimes{\cal L}\lambda.
\]
This implies that
\begin{displaymath}
  \E^P\left\{W(z,\bx_1,\bk_1)W(z,\bx_2,\bk_2)\right\}=
\E^P\left\{W(z,\bx_1,\bk_1)\right\}\E^P\left\{W(z,\bx_2,\bk_2)\right\}
\end{displaymath}
by uniqueness of the solution to the above transport equation with 
initial conditions given by $W_0(\bx_1,\bk_1)W_0(\bx_2,\bk_2)$. This 
proves that the limiting measure ${P}$ is deterministic and unique
(because characterized by the transport equation) and that the sequence
$W_\eps(z,\bx,\bk)$ converges in probability to $W(z,\bx,\bk)$.

\end{document}